\documentclass[conference]{IEEEtran}
\usepackage{times}

\usepackage[numbers]{natbib}
\usepackage{multicol}
\usepackage[bookmarks=true]{hyperref}
\usepackage[
left=0.625in,right=0.625in,top=0.75in,bottom=1in,%
footskip=.1in]{geometry}
\usepackage{siunitx}
\usepackage{footmisc}
\usepackage[T1]{fontenc}
\usepackage[utf8]{inputenc}
\usepackage{authblk}
\usepackage{xcolor}
\usepackage{color}
\usepackage{subfig}
\usepackage{graphics} 
\usepackage{graphicx} 
\usepackage{amsmath} 
\usepackage{makecell} 
\usepackage{multirow} 
\usepackage{textcomp}

\newcommand{\Skip}[1] {}

\newcommand{\YOON}[1] {}
\newcommand{\IK}[1] {}
\newcommand{\Doheon}[1] {}

\newcommand{\JW}[1] {}

\begin{document}

\title{Robust Sound Source Localization considering Similarity of Back-Propagation Signals}


\author[1]{Inkyu An}
\author[1]{Doheon Lee}
\author[2]{Byeongho Jo}
\author[2]{Jung-Woo Choi}
\author[1]{Sung-Eui Yoon}
\affil[1]{School of Computing}
\affil[2]{School of Electrical Engineering}
\affil[ ]{Korea Advanced Institute of Science and Technology}
\affil[ ]{Daejeon, South Korea}
\affil[ ]{Email: \{inkyu.an, doheonlee, byeongho, jwoo\}@kaist.ac.kr, sungeui@kaist.edu}

\renewcommand\Authands{ and }





%

\maketitle

\begin{abstract}
We present a novel, robust sound source localization algorithm considering
	back-propagation signals.  Sound propagation paths are estimated by
	generating direct and reflection acoustic rays based on ray tracing in
	a backward manner.  We then compute the back-propagation signals by
	designing and using the impulse response of the backward sound
	propagation based on the acoustic ray paths.  For identifying the 3D
	source position, we suggest a localization method based on the
	Monte Carlo localization algorithm.  Candidates for a source
	position is determined by identifying the convergence regions of
	acoustic ray paths.  This candidate 
	is validated
	by measuring similarities between back-propagation signals, under the assumption that
	the back-propagation signals of different acoustic ray paths should be
	similar near the sound source position.  Thanks to considering
	similarities of back-propagation signals, our approach can localize a
	source position with an averaged error of 0.51~m in a room of 7~m by 7~m area with 3~m height
	in tested environments.  We also observe 65~\% to 220~\% improvement
in accuracy over the state-of-the-art method.  This improvement 
is achieved in environments containing a moving source, an
obstacle, and noises.
\end{abstract}

\IEEEpeerreviewmaketitle

\section{INTRODUCTION}
\label{sec:introduction}

As robots become more widely available, it is getting more imperative for a
robot to understand environments for safe and accurate operations.
There have been many kinds of research efforts to perceive the environments by
acquiring and using data from hardware sensors.
One of main the research topics for understanding the environments focus on
identifying locations of a robot itself and other objects in environments from
collected data by vision cameras and depth sensors.
Departing from these approaches, an acoustic data measured by acoustic sensors
has recently attracted attention as an important clue for localizing various
objects.

The problem identifying the location of a sound source  from collected acoustic
data is widely known as the sound source localization (SSL).  There have been
many types of SSL researches 
including the use of a time difference of arrival (TDOA) of sound waves at
microphones~\cite{knapp1976generalized,valin2007robust} and 
advanced methods using a spherical microphone array based on the analysis of
the spherical harmonics functions~\cite{khaykin2009coherent, li2011spherical,
rafaely2005phase,	yan2011optimal}.

These approaches, unfortunately, are not designed mainly for the estimation of  3D position but for the identification of an incoming direction of sound.
Especially, when a sound source is occluded by an obstacle or
multiple sound sources are located in the same direction, most of prior approaches
cannot specify the location of the source generating the sound signal.
To address this issue, recent techniques were proposed to find a 3D source
location even if the sound source is in the non-line-of-sight
state~\cite{an2018diffraction, an2018reflection}.
These techniques estimate sound propagation paths from the source to
microphones as acoustic rays, generated by the ray tracing technique, and
identify the 3D source location by using generated acoustic rays.
However, the accuracy of these methods decreases in noisy environments
including a moving sound and obstacles.

\begin{figure}[t]
	\centering
	\includegraphics[width=0.9\columnwidth]{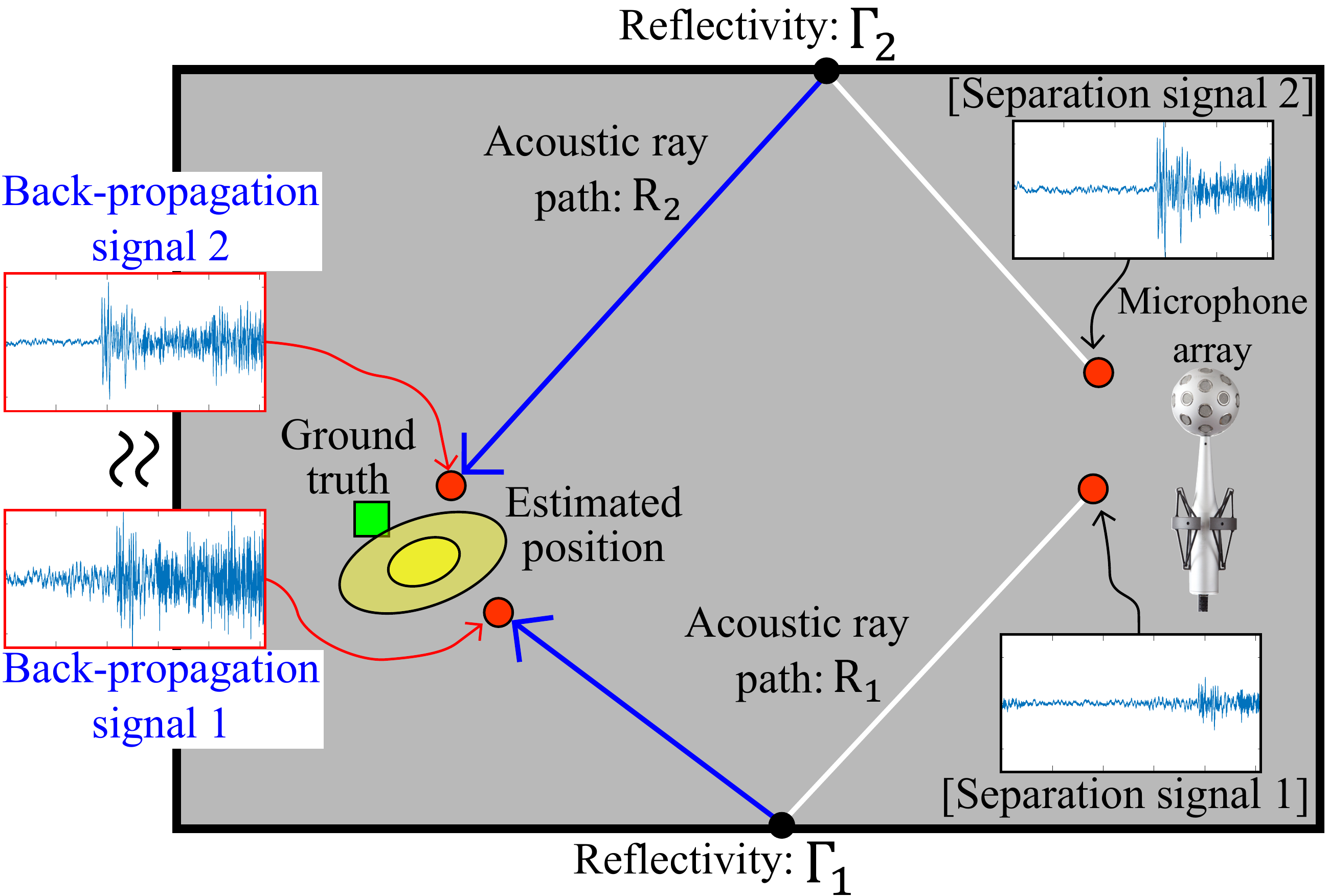}
	\caption{
		Our approach generates direct and indirect acoustic ray paths
		and localizes the sound source, while considering
		back-propagation signals on generated acoustic ray paths.  The
back-propagation signals are virtually computed signals that could be heard at
particular locations and computed by using impulse responses, considering the
travel distance and reflection
		amplification of acoustic ray paths from separation signals
		that are extracted from measured signals at the
		microphones. When two back-propagation signals of acoustic ray
		paths are highly correlated, we treat them to be originated
		from the same source.
		\YOON{change the color of the ref. text to black; it was emphasized. I would emphasize the color of back-propagation by using blue text} 
	}
	\vspace{-1.0em}
	\label{fig:summary}
\end{figure}


\Skip{
We aim to improve the estimation performance by computing back-propagation
signals that could be heard at a particular point.  The propagation paths of a
sound are first estimated by generating the acoustic ray path based on a ray
tracing technique in a backward manner.  We then compute the back-propagation
signals, traveled to the reverse direction of the propagation path, using the
impulse responses of acoustic ray paths.
When acoustic ray paths are from different sound sources, their
back-propagation signals at a converging point are highly likely to be
incoherent. In this regard, we consider the similarities of
back-propagation
signals for robustly identifying the source location.
}

\textbf{Main Contributions.}
To robustly identify the sound source location, we present a novel, sound
source localization algorithm using back-propagation signals (Fig.~\ref{fig:summary}).
Using a beamforming algorithm, we first compute incoming directions of the
sound and separation signals corresponding to those specific incoming
directions (Sec.~\ref{sec:beamforming}).
We then estimate sound propagation paths by generating  acoustic ray paths in
the reverse direction to the incoming directions of the sound
(Sec.~\ref{sec:acoustic_ray_tracing}), and compute the back-propagation signals
using the impulse response of the acoustic ray path from the separation signal
(Sec.~\ref{sec:back_propagated_signals}).
Intuitively speaking, back-propagation signals are virtually computed signals
that could be heard at a particular location in acoustic paths from the
measured signals at the microphone array.

Finally, we  use the Monte Carlo localization algorithm estimating a location
of the sound as a converging region of computed acoustic ray paths.  In
particular, we utilize the computed back-propagation signals of different
acoustic ray paths
for accurate estimation of the sound location, under the intuitive
assumption that acoustic paths coming from the same sound source should have
similar back-propagation signals at the estimated location
(Sec.~\ref{sec:estimating_source_position}).

Given environments containing a moving source, an obstacle, and noises in a 7m$\times$7m$\times$3m room, our localization algorithm estimates the location of the sound source with an error of 0.5m on average.
The localization algorithm of our approach is improved by 65\%\textendash 220\% compared
to the prior work that does not consider the back-propagation signals.

~

~\Skip{
Recently, there are many efforts to find a 3D source position by approximating the sound propagation paths using acoustic rays~\cite{an2018reflection,an2018diffraction}.
Their methods localize the sound source via identifying the convergence of acoustic rays based on the hypothesis that every path of sound propagation is came from a same source.
However, because the convergence regi        ons occur at many locations in complex environments containing a dynamic sound source, obstacles, and noises, the accuracy might decrease.
}

\section{RELATED WORKS}

In this section, we give a brief overview of prior works on sound source localization and sound propagation. 

\subsection{Sound source localization}
There has been a significant amount of efforts to localize a sound source by
estimating a direction of sound.  Many works have been studied based on time
difference of arrival
(TDOA) of a microphone array.  
Knapp \textit{et al.}~\cite{knapp1976generalized} suggested an efficient algorithm to estimate a time delay of sound arrivals at each microphone pair.  

When the shape of the microphone array is spherical, 
one can uniformly estimate angle of sound sources in every azimuth and elevation angle.
Therefore, many beamformer algorithms
have been studied by using the spherical Fourier transform, which
transforms the function on the unit sphere to the
spherical harmonics domain.
Rafaely~\cite{rafaely2005analysis,rafaely2015fundamentals} presented a
theoretical framework of analysis of spherical microphone array.
This approach localized the source direction by computing the beam energy map
on the unit sphere using steered beamformer algorithms.  
Valin \textit{et al.} \cite{valin2007robust} suggested a memoryless
localization algorithm based on a simple delay-and-sum beamformer on the space
domain using 8 microphones located on the surface of the sphere, 
and Rafaely \cite{rafaely2005phase} extended the delay-and-sum localization method to process on the spherical harmonics domain.
Yan \textit{et al.}~\cite{yan2011optimal} suggested a localization algorithm
using the minimum variance distortionless response (MVDR) power spectra on the
spherical harmonics domain.
Li \textit{et al.}~\cite{li2011spherical} presented a MUSIC (Multiple Signal
Classification) based beamformer algorithm, which uses an orthogonality between
a noise-only subspace and a signal-plus-noise subspace on the spherical
harmonics domain.
Khaykin \textit{et al.}~\cite{khaykin2009coherent} presented 
a frequency smoothing technique 
for spherical microphone arrays.
Unfortunately, these techniques were designed for detecting incoming
directions, not the 3D location of a sound source in an arbitrary environment.

Recently, 3D sound source localization methods have emerged by considering not
only the direct path, but also indirect paths such as reflection and diffraction of
sound propagation.  An \textit{et al.}~\cite{an2018reflection} suggested a
reflection-aware SSL algorithm via approximating sound propagation as acoustic
rays generated by a ray tracing technique.  
This technique was extended to 
a diffraction-aware localization method for handling a non-line-of-sight sound
source~\cite{an2018diffraction}.
These methods localize the 3D source position by identifying the convergence
region of acoustic rays based on the hypothesis that acoustic rays are
generated from the same source.
However, if there exist many convergence regions of acoustic rays in 
environments containing a moving sound, obstacles, and noises, we observe that
the accuracy deteriorates (Sec.~\ref{sec:result_and_discussion}).
In this paper, we aim to overcome this issue 
by utilizing the acoustic signals back-propagated to the source location.
\Skip{
and utilizing a spherical Fourier
transform based
beamforming method.
}


\begin{figure*}[t]
	\centering
	\includegraphics[width=2\columnwidth]{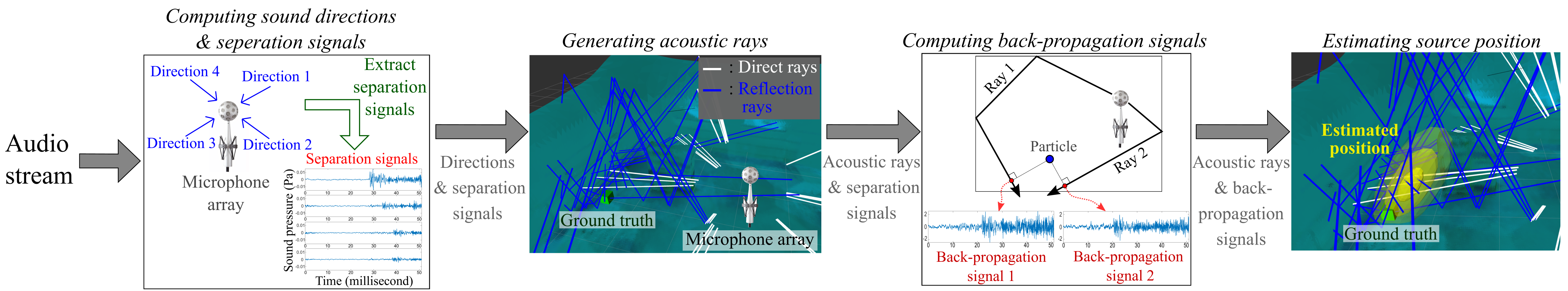}
	\caption{
		An overview of the proposed method. Detailed explanations are in Sec.~\ref{sec_overview}.
	}
	\vspace{-1.0em}
	\label{fig:overview}
\end{figure*}

\subsection{Sound propagation}

For the generation of acoustic rays and back-propagation signals, an accurate sound propagation model should be employed.
Various sound propagation models have been studied for generating a realistic sound in a
virtual environment.  For generating a realistic sound, these prior methods
have focused on modeling how the sound emitted at the source is propagated to
the listener.
\Skip{
We want to restore the initial sound signal occurred at the source based on the
techniques of the sound propagation simulations and utilize the restored
signals to improve the accuracy of the sound source localization algorithm.
}

At a broad level, sound propagation techniques can be categorized as numerical
acoustic (NA) and geometric acoustic (GA) methods.  NA techniques have been
studied based on modeling the sound propagation using the acoustic wave
equation.
Since it requires considerable amount of computation time to solve the acoustic wave equation,
there are difficulties in extending NA techniques to real-time applications.

GA techniques are based on ray tracing algorithms and facilitate an efficient
sound simulation for real-time applications by estimating an acoustic impulse response between a source
and a listener in an environment.  
 Cao \textit{et al.}~\cite{cao2016} enabled efficient
computation of sound simulation through bidirectional path tracing.  Li
\textit{et al.}~\cite{li2018scene} proposed an efficient sound simulator for
\ang{360} videos by handling early and late reverberation separately.

There have been hybrid approaches supporting various low-frequency wave
properties within GA methods.  Yeh \textit{et al.}~\cite{yeh2013wave} presented
a hybrid approach that combines geometric and numeric methods for handling
complex environments.  Schissler \textit{et al.}~\cite{schissler2014high}
suggested an efficient way to deal with reverberation by considering
high-order diffraction and diffuse reflection in large environments. 

For improving the accuracy of sound source location, which is an inverse
problem to the sound generation, we adopt the concept of an impulse response
used for these sound generation methods.  The main difference
over these sound generation method is that we need to reversely propagate
signals measured at the microphone array to ones that can be heard in
particular locations on acoustic paths.

\section{OVERVIEW}
\label{sec_overview}

\Skip{
We propose a novel sound source localization algorithm that utilizes calculated signals, 
which are propagated along acoustic rays from measured audio signals at microphones; we call the calculated signals \YOON{which one?}\IK{Addressed.} as the back-propagated signal. \Doheon{I think this paragraph describes inverse-propagation signal, but explanation is not quite interpretable. Sound simulation from measure point to inverse received direction?}\IK{Addressed.}
}

Our sound source localization algorithm utilizes signals, called
back-propagation signals,  that are back-propagated to particular locations on
sound propagation paths from signals measured at the microphone.

The overview of our algorithm is shown in Fig.~\ref{fig:overview}.  
Our method expresses a surrounding environment in form of a mesh map,
 which is reconstructed from the point cloud collected by the depth sensor.  
At runtime, audio streams
are collected by a 32 channel microphone array.  After localizing incoming
directions of sound using a MVDR (minimum variance distortionless response)
based beamformer algorithm, we estimate acoustic signals observed from major
incoming directions by applying beam patterns~\cite{rafaely2015fundamentals}.

For simulating acoustic paths, we generate direct and reflection acoustic rays
by applying ray tracing in the backward manner~\cite{an2018reflection}.
Specifically, 
we generate direct acoustic rays in the opposite directions to those of incoming sounds.
Once these direct acoustic rays intersect with the surrounding environment, 
we generate reflection acoustic rays to reversely
simulate the reflection effect.

Finally, we perform the Monte Carlo localization algorithm for identifying a source
position from the generated acoustic paths.
If these acoustic rays are actually coming from
the same sound source, back-propagation signals 
at a candidate location should be similar to each other.
We therefore utilize 
those back-propagation signals of acoustic rays at a candidate location as an important factor of identifying the sound source location.
This back-propagation signal of an acoustic path is computed by the impulse
response that is initialized with the separation signal estimated for each
direct acoustic ray.

\Skip{
the convergence of acoustic rays as an important factor for localizing the sound source.
However, because the accuracy of the localization method, which is the
reflection-aware sound source localization algorithm~\cite{an2018reflection},
only using convergence of acoustic rays decreases in complex environments
(Fig.~\ref{fig:graph_w_obs} and Fig.~\ref{fig:graph_wo_obs}), 
we try to verify whether the convergence region of acoustic rays is close to the actual sound source using back-propagated audio signals.
Ideally, the back-propagated signals, which is restored to the propagated signal along the propagation path of sound, should be the same with the initial sound signal emitted at the source.
}

\Skip{
We compute the back-propagated signals near the convergence region \YOON{you
didn't define it}\IK{Addressed} of acoustic rays using the impulse response of
the acoustic ray.  

back-propagated signals are used for validating
whether the convergence region of acoustic rays is close to the actual source
position.
}

\section{Sound source localization using back-propagated signals}

In this section, we describe each module of our approach illustrated in
Fig.~\ref{fig:overview}.

\subsection{Beamforming}
\label{sec:beamforming}

In a real environment involving moving sound sources, obstacles, or noise,
acoustic rays generated naively by our approach may converge to a position
other than the actual location of the sound source.
We also found that this occurs in practice and thus its accuracy decreases
in previous works~\cite{an2018reflection}.  To solve this
problem, we aim to generate and utilize back-propagation signals to a candidate
3D location along acoustic rays.  This back-propagation signals at a location
can be computed by simulating the reverse process of sound propagation, 
i.e., by reversely performing ray tracing.

The input signals measured at the microphone 
consist of many different signals that were
propagated through different paths from a sound source. Ideally, we want to
estimate the propagation paths of those signals using acoustic rays and restore
back-propagation signals on a particular position on those propagation paths.

To generate acoustic rays, we estimate incoming sound directions at the
microphone using a beamforming algorithm~\cite{rafaely2015fundamentals}.  Note
that our
input signals are measured at discrete locations of the microphones, 
but each microphone signal is, in fact, a mixture of signals from different directions.
We therefore aim to compute signals along incoming
directions, and use the beamforming algorithm.

\begin{figure}[t]
	\centering
	\includegraphics[width=0.85\columnwidth]{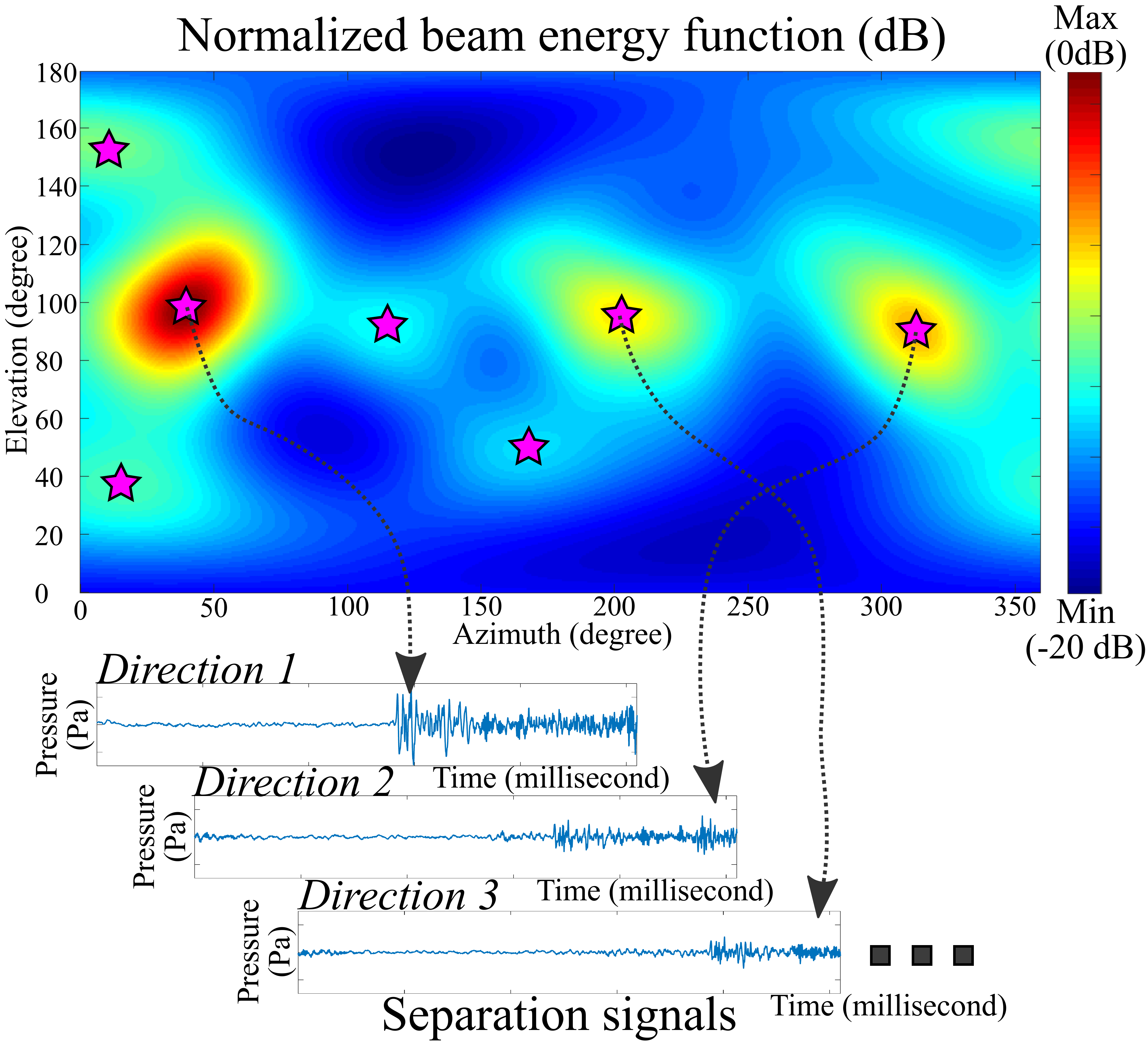}
	\caption{
		A beam energy function computed by a 
		beamforming algorithm, where the horizontal axis is the azimuth
		angle and the vertical axis is the elevation angle of the unit sphere.  Local
		maxima of the beam energy function are treated most significant incoming directions
		of sound.  The separation signal of each incoming direction
		is extracted by using the beam pattern from the input signals
		measured by microphones.  }
	\vspace{-1.0em}
	\label{fig:compute_separation_signals}
\end{figure}

Fig.~\ref{fig:compute_separation_signals} shows a beam energy function 
representing a magnitude distribution of sound signals on the unit
sphere computed by the beamforming method.
We then extract sound signals incoming from the directions of dominant magnitude by applying beam patterns~\cite{rafaely2015fundamentals} steered to those directions.

We utilize the minimum variance distortionless reponse (MVDR) based beamforming
algorithm ~\cite{rafaely2015fundamentals, yan2011optimal} for computing
incoming directions of the sound.  
Other high-resolution beamforming techniques, such as MUSIC~\cite{li2011spherical}, can also be applied in here, 
but MUSIC based beamformers showed rather inconsistent results depending on the number of assumed sources.
Fig.~\ref{fig:compute_separation_signals} also shows local maxima of the beam
energy function $y(\theta, \phi)$, computed by the MVDR, on
the unit sphere representing 
the incoming directions of the sound:
\begin{equation}
	[\hat{d}_1, \hat{d}_2, \cdots, \hat{d}_N]=f_{max}\{ y(\theta, \phi) \},
	\label{eq:compute_arrival_directions}
\end{equation}
where $\hat{d}_n$ denotes a directional vector of the $n$-th local maximum on
the unit sphere among $N$ different local maxima in a frame,
$\theta$ is an elevation angle, $\phi$ is an azimuth angle, and
$f_{max}\{\cdot\}$ is a function for finding local maxima of the beam energy
function.
In practice, we identify 8 local maxima on average in our tested experiments.

The computed directional vectors $[\hat{d}_1, \hat{d}_2, \cdots, \hat{d}_N]$ are
used as a set of inverse directions of incoming sounds, and we thus generate our acoustic
rays in those estimated directions, to
simulate the back-propagation of sound paths.
We then extract
separation signals, which could be heard in those incoming directions.
For the $n$-th direction $\hat{d}_n$, the separation signal $S_n[f]$ is computed
by designing and using a beamforming weight $W_n[f]$, which is a beam pattern in
the spherical harmonic domain~\cite{rafaely2015fundamentals}:
\begin{equation}
	S_n[f] = M[f] \cdot \{ W_n[f] \}^{*},
	\label{eq:extract_separation_signals}
\end{equation}
where $f$ is a frequency, $M$ is the spherical harmonic coefficients, which are
measurement signals (32 channels) transformed by spherical Fourier transform, 
and
$\{\cdot\}^*$ is the complex
conjugate.
All of variables, $S_n[f]$, $M[f]$, and $W_n[f]$
contain data for $L$ frequency bins ranging 
from 0 to 24 kHz.

\subsection{Acoustic ray tracing}
\label{sec:acoustic_ray_tracing}

We explain how to generate acoustic rays from estimated directions
$[\hat{d}_1, \hat{d}_2, \cdots, \hat{d}_N]$ that are the inverse directions of incoming sounds.
We want to estimate propagated paths 
(e.g., direct and reflection path) 
of the sound from its source location to the microphone
array location using the acoustic rays.  We generate such acoustic rays
considering direct and reflection paths based on the RA-SSL
algorithm~\cite{an2018reflection}.

\begin{figure}[t]
	\centering
	\includegraphics[width=0.8\columnwidth]{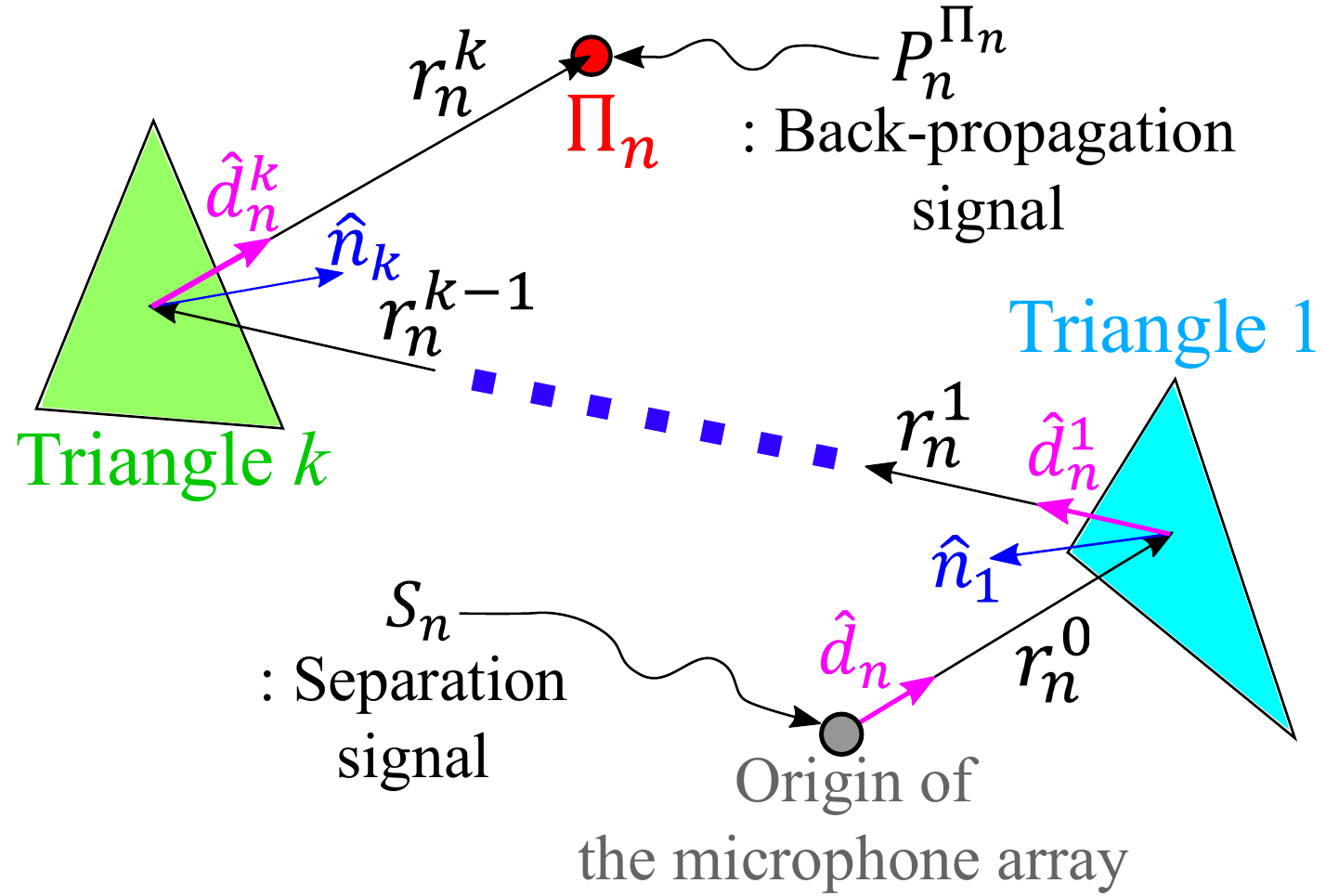}
	\caption{
		An example of generating an acoustic ray
		path $R_n$ and its back-propagation signal.  The primary
		acoustic ray, $r_n^0$, of the $n$-th acoustic ray path $R_n$ is generated
		to the direction vector $\hat{d}_n$ that is the inverse direction of the $n$-th incoming sound.  When the
		acoustic ray $r_n^0$ hits an obstacle represented by Triangle
		1, its reflection acoustic ray $r_n^1$ is generated according
		to the specular reflection based on the
		normal vector $\hat{n}_1$ of Triangle 1.  The
		back-propagation signal $P_n$ is computed by using the impulse
		response of $R_n$ at a specific point, $\Pi_n$, on the path
		from the separation signal $S_n$. 
	}
	\vspace{-1.0em}
	\label{fig:generate_reflection_ray}
\end{figure}

For the $n$-th acoustic ray path, denoted by $R_n$, its primary acoustic ray,
$r_n^0$, is created in to the $n$-th direction vector $\hat{d}_n$, as shown in
Fig.~\ref{fig:generate_reflection_ray}.  If the acoustic ray collides with an
obstacle, its secondary, reflection ray is generated by assuming the specular
reflection, and is denoted by $r_n^1$, where the superscript represents the
order of the acoustic ray path.  When $R_n$ is propagated until a $K$-th order,
the acoustic ray
path $R_n$ consists of $K$ acoustic rays: i.e., $R_n =
[r_n^0, r_n^1, \cdots, r_n^{K-1}]$.

\subsection{Back-propagation signals}
\label{sec:back_propagated_signals}

We introduce how to compute back-propagation signals based on acoustic ray
paths $[R_1, R_2, \cdots, R_N]$ and separation signals $[S_1, S_2, \cdots,
S_N]$; there is a tuple of  $(R_n, S_n)$ for the direction vector
$\hat{d}_n$ that is the inverse direction of the $n$-th incoming sound.  We want to compute the back-propagation signal $P_n$ from the
separation signal $S_n$ by designing and using an impulse response of backward
sound propagation based on the acoustic ray path $R_n$.  The impulse response
describes the reaction of any linear system as a function of time-independent
variables; the input is the separation signal and the output is the
back-propagation signal in our system.

\Skip{
There have been many efforts to model the sound propagation using the impulse
response in virtual environments for generating realistic
sound~\cite{schissler2014high, yeh2013wave, li2018scene}.  They estimate the
impulse response between the sound source and the listener by generating a huge
amount of rays.  If the sound source and listener are connected by a path of
acoustic rays, they compute the impulse response of the connected ray path.  By
adding all connected impulse responses between the source and listener, they
estimate the whole impulse response of the sound propagation system that is a
forward simulation.  
}

In this work, we utilize the impulse response for the backward propagation to
improve the accuracy of the sound source localization.
In the forward sound propagation, the impulse response of an acoustic ray path
is described by attenuations according to the travel distance of a ray path and
reflection.
For example, the travel distance attenuation represents the decrease of sound
pressure inversely proportional to the travel
distance of the ray path, because the sound is propagated according to the
spherical wave in 3D environments; similar for the reflection attenuation.
\Skip{
	The reflection attenuation represents the
decrease of pressure of the reflected sound on the obstacle, since the sound
energy is absorbed on the obstacle surface based on the material
characteristic.  
}

On the other hand, for the backward propagation problem, the attenuation of
travel distance and reflection becomes an amplification of the sound pressure.
Suppose that we aim to  compute the back-propagation signal from the starting
point to a specific point $\Pi_n$ (Fig.~\ref{fig:generate_reflection_ray}) on
an acoustic ray path using the backward impulse response, where there is the
$n$-th tuple $(R_n, S_n)$ and the acoustic ray path $R_n$ consists of $K$
acoustic rays $[r_n^0, \cdots, r_n^{K-1}]$;
$r_n^{0}$ is a primary ray and $r_n^{k}$ is the $k$-th reflection ray $(1\le k
\le K-1)$.  In the frequency domain, the backward impulse response
$H_n^{\Pi_n}$ is
described by amplifications because of the travel distance $l$ and the
reflection until the $k$-th order reflection ray $r_n^k$:
\begin{equation}
	H_n^{\Pi_n}[f] = \exp \left( \frac{j2\pi fl}{c} \right) \cdot A^{D}[l] \cdot A^{R}[R_n, k, f],
	\label{eq:compute_backward_impulse_response}
\end{equation}
where the term inside the exponential function is 
for shifting the back-propagation signal to the time delay of the sound
propagation at the specific point $\Pi_n$
and $c$ is the speed of sound.
$A^{D}$ 
is a coefficient of the travel
distance amplification, and is defined 
by a function of the travel distance $l$:
$A^{D}[l] = 4 \pi (1+l)$. 
Also, $A^{R}$ is a
coefficient of the reflection amplification,
and is defined
by considering specular reflections until the $k$-th order reflection ray:
\begin{equation}
	A^{R}[R_n, k, f] = \prod_{i=1}^{k}{\left[\frac{1}{\Gamma_i[f]}\right]},
	\label{eq:compute_reflection_amplification}
\end{equation}
where $\Gamma_i$ denotes the reflectivity (reflection coefficient) of the triangle hit by the $(i-1)$-th order ray; 
the reflection coefficient is a function of frequency $f$ and we refer to
coefficient values reported by \cite{schissler2018acoustic}.  

The back-propagation signal $P_n^{\Pi_n}$ at the specific point $\Pi_n$ on
the acoustic ray path $R_n$ is finally computed by the product of the backward
impulse response $H_n^{\Pi_n}$ and the separation signal $S_n$ in the
frequency
domain: 
\begin{equation}
P_n^{\Pi_n}[f] = S_n[f] \cdot H_n^{\Pi_n}[f].
\label{eq:compute_back_propagation_signal}
\end{equation}

\subsection{Estimating a source position}
\label{sec:estimating_source_position}

We now explain how to estimate the source position using back-propagation
signals of acoustic ray paths. 

\Skip{
	Our approach for localizing the source is based
on two hypotheses: 1) when there is a single sound source, acoustic ray paths
should converge at the source location and 2) back-propagation signals should
be similar near the location.  
Acoustic ray paths are propagated close to the source using the backward
acoustic ray tracing algorithm based on different types of propagation paths.
}

The Monte Carlo sound source localization algorithm identifying the
convergence region of acoustic ray paths was suggested in the prior work
(RA-SSL)~\cite{an2018reflection}; the convergence region means the area where acoustic ray paths gather.
However, in real environments containing
a moving source, 
obstacles,
and noises, the accuracy of RA-SSL  can decrease.  
For example, when there are background noises or complex configurations with
obstacles, they can trigger to generate many arbitrary or incoherent acoustic
ray paths, which may cause many convergence regions. 
If some cases, those convergence regions may be considered stronger than the one of the ground truth, deteriorating the accuracy of the localization method.
\Skip{
if the convergence of acoustic ray paths near the ground truth is weaker than
any other regions because of a moving source and an obstacle that cause
high-order reflection acoustic rays, their algorithm estimates the source
position in another region where more acoustic ray paths gather. 
}
By considering back-propagation signals, we aim to identify those arbitrary and
incoherent acoustic ray paths and cull away acoustic ray paths with different
back-propagation signals indicating that they are from different sound sources.
\Skip{
address the limitation of
the localization algorithm that is based on identifying the convergence region
of ray paths.
}

Intuitively speaking, if there are two acoustic ray paths caused by the same
source, their back-propagation signals should be similar near the location of
their sound source.  In other words, when back-propagation signals of two
acoustic ray paths are different at a location, the location is unlikely to be
a candidate for a converging region of the sound source.
Based on this observation, we design and utilize similarity between
back-propagation signals for robustly identifying the sound source's location
even in these difficult settings.

\begin{figure}[t]
	\centering
	\includegraphics[width=0.7\columnwidth]{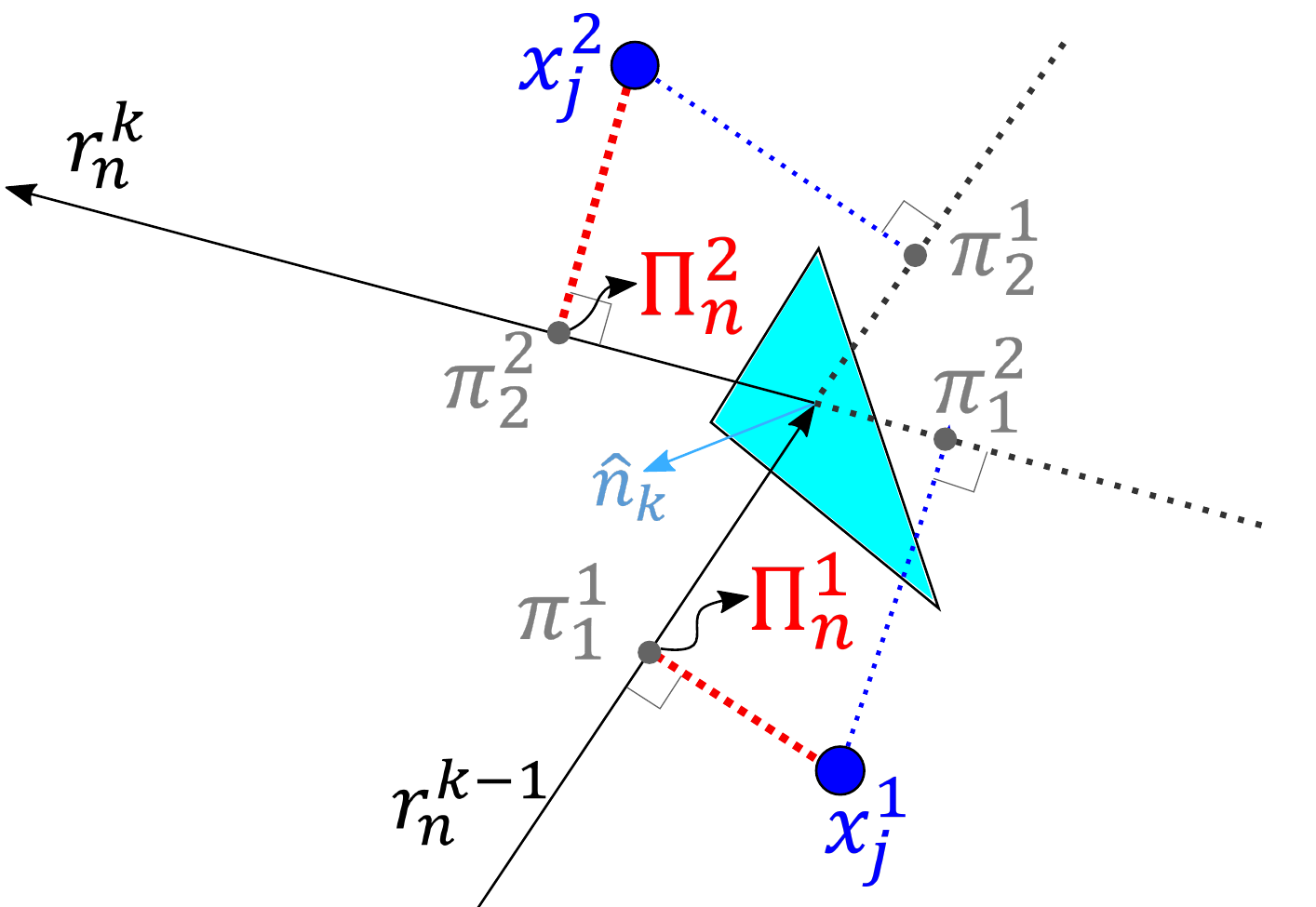}
	\caption{
		Examples of determining the point of the
		acoustic ray path for computing the back-propagation signal.
		For the particle of $x^2_j$, the perpendicular foots $\pi_2^k$
		on all $k$-th order acoustic rays of the $n$-th acoustic ray path are
		computed.  We then decide the representative perpendicular foot
		$\Pi_n^2$ satisfying the shortest distance from $x_j^2$ to
		$R_n$.}
	\vspace{-1.0em}
	\label{fig:distance_weight}
\end{figure}

Our estimation process is based on the Monte Carlo localization algorithm
consisting of three parts: sampling, computing a weight of particles, and
resampling.
The main
differentiation of our approach over the prior technique is that our method
improves the localization accuracy based on  a novel module
for computing weights of particles based on our back-propagation signals.

Suppose there are $i$-th particles, $x_j^i$, representing hypothetical
locations of the sound source at a $j$ frame. 
We would like to compute 
how close the particle is to acoustic ray paths.
For this, we define a specific point $\Pi_n^i$, which is decided to be the
point  satisfying the shortest distance between $x_j^i$ and any point on the
$n$-th acoustic ray path;
i.e., $\Pi_n^i = \textrm{argmin}_{\pi_i^k}{\vert\vert x_j^i-\pi_i^k \vert\vert}$, where
$\pi_i^k$ is the perpendicular foot on the $n$-th acoustic ray path from the
$x_j^i$ position (Fig.~\ref{fig:distance_weight}).  We then compute our
back-propagation signal according to
Eq.~\ref{eq:compute_backward_impulse_response} on the shortest point $\Pi_n^i$
on the $n$-th acoustic ray path from the particle $x_j^i$.

From the back-propagation signal $P_n^{\Pi_n^i}[f]$ in the frequency domain,
we compute the back-propagation signal $p_n^{\Pi_n^i}[t]$ in the time domain
signal.
We then calculate a particle weight, $w_j^i$, representing the probability of
being a convergence region of the sound source,  based on two factors: a
distance weight, $w_d$, representing how away the particle is from
the $n$-th acoustic ray path \YOON{check}
and a similarity weight, $w_s$, indicating how similar between
$p_n^{\Pi_n^i}[t]$ and
other signals given acoustic ray paths:
\begin{equation}
	w_j^i = P(O_j \vert x_j^i) = \frac{1}{n_c} \sum_{n=1}^{N_j}{[ w_d(x_j^i, R_n) + \alpha \cdot w_s(x_j^i, R_n) ]},
	\label{eq:compute_particle_weight}
\end{equation}
where $N_j$ is the number of acoustic ray paths at the $j$ frame, $O_j$ is the
observation containing $[P_1^{\Pi_1^i}, \cdots, P_{N_j}^{\Pi_{N_j}^i}]$ and
$[R_1, \cdots, R_{N_j}]$, $n_c$ is a normalizing constant, and $\alpha$
denotes a parameter for adjusting different weights.  

The distance weight $w_d$ is calculated by using the Euclidean distance between
$x_j^i$ the particle location and the point $\Pi_n^i$:
\begin{equation}
w_d(x_j^i, R_n) = G (\vert\vert x_j^i - \Pi_n^i \vert\vert \, \vert  0, \sigma_w),
\label{eq:compute_distance_weight}
\end{equation}
where $G$ is the Gaussian distribution
function with the zero mean and a standard deviation $\sigma_w$.
$w_d$ is
maximized when the particle $x_j^i$ is on the perpendicular foot $\Pi_n^i$,
which is on the $n$-th acoustic ray path \YOON{check}.
The similarity weight $w_s$  measures the similarity between the
back-propagation signal $p_n^{\Pi_n^i}$ 
from the $n$-th acoustic ray
path and ones of other acoustic ray paths:
\begin{equation}
	w_s(x_j^i, R_n) = \frac{1}{n_s} \sum_{\substack{m=1, \\ m \ne n}}^{N_j}
	\begin{cases}
    	\frac{L-l_{cc}(n, m)}{L},& \text{if } a_{cc} (n,m) > a_{th}\\
		0,              & \text{otherwise,}
	\end{cases}
\label{eq:compute_similarity_weight}
\end{equation}
where $n_s$ is the normalizing constant, $L$ is the length of the back-propagation signal, 
$a_{cc}(\cdot)$ is the peak coefficient in a normalized range of $-1$ to $1$, $l_{cc}(\cdot)$ is the peak
coefficient delay, and $a_{th}$ denotes the threshold value of $a_{cc}(\cdot)$.
Both variables of $a_{cc}(\cdot)$ and $l_{cc}(\cdot)$
are computed by
applying the
cross-correlation operation between two signals, $n$-th and $m$-th signals:
\begin{equation}
\begin{aligned}
	a_{cc} (n, m) &= \textrm{max} \{ (p_n^{\Pi_n^i} \star p_m^{\Pi_m^i})[\tau] \}, \\
	l_{cc} (n, m) &= \textrm{argmax}_{\tau} \{ (p_n^{\Pi_n^i} \star p_m^{\Pi_m^i})[\tau] \}, 	
\end{aligned}
\label{eq:compute_cross_correlation}
\end{equation}
where $\star$ is the cross-correlation operator.

\begin{figure}[t]
	\centering
	\includegraphics[width=0.85\columnwidth]{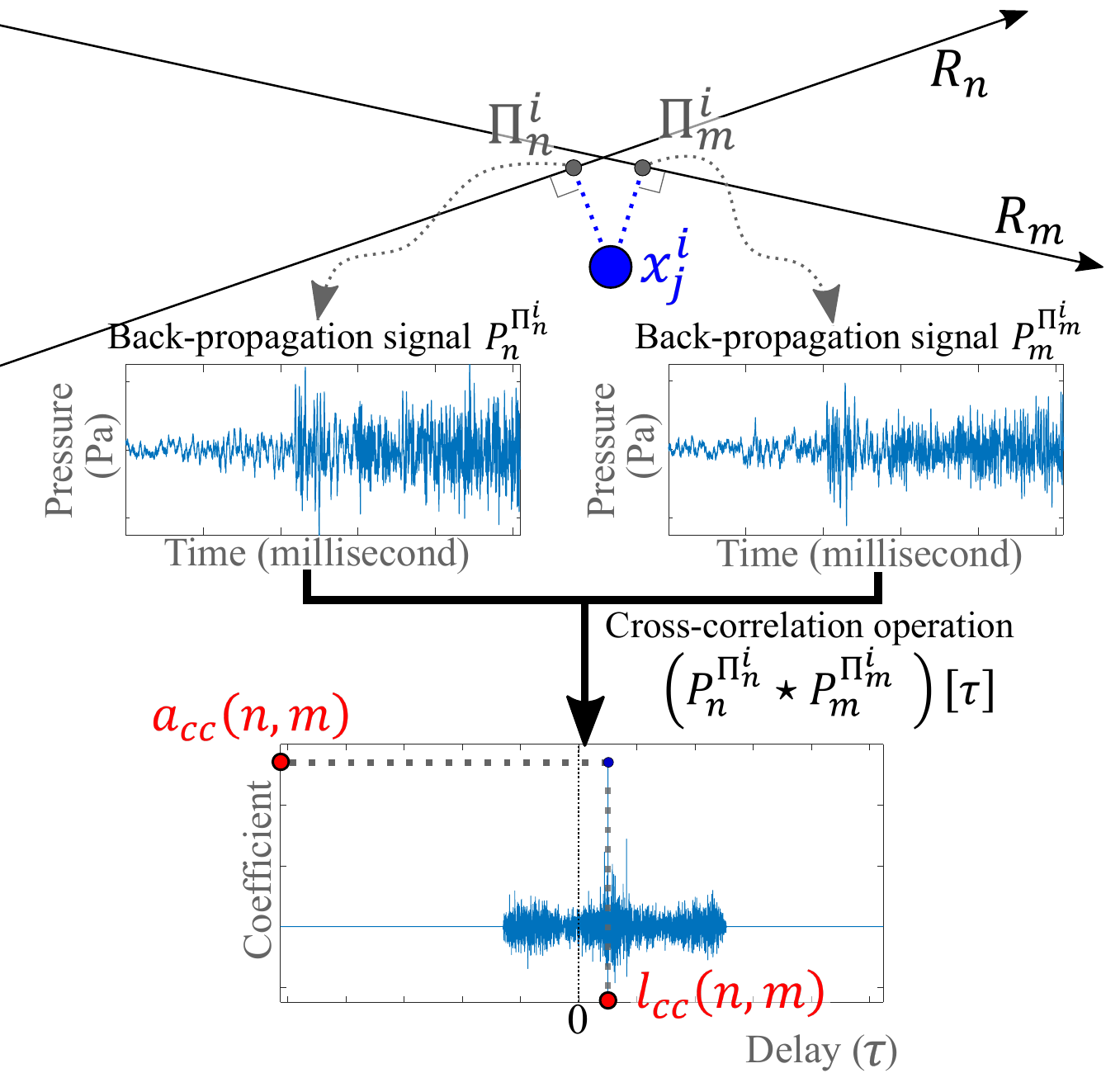}
	\caption{
		An example of computing the peak coefficient
		$a_{cc}$ and the peak coefficient delay $l_{cc}$ by using the
		cross-correlation operation.  
		Given two back-propagation signals, $p_n^{\Pi_n^i}$ and
		$p_m^{\Pi_m^i}$ at $\Pi_n^i$ and $\Pi_m^i$, respectively,
		we perform 
	the cross-correlation operation between two signals.
		The maximum coefficient becomes the peak coefficient $a_{cc}$
		and 
	the time delay from the time origin, 0, to the time realizing 
	the
	maximum coefficient becomes the peak coefficient delay
		$l_{cc}$.
	}
	\vspace{-1.0em}
	\label{fig:similarity_weight}
\end{figure}

As shown in Fig.~\ref{fig:similarity_weight}, $a_{cc}(\cdot)$ represents how
much both back-propagation signals are correlated, and $l_{cc}$ shows the time
difference of occurrence between both back-propagation signals. As both
back-propagation
singles are from the same sound source, ideally $a_{cc}$ and $l_{cc}$ become
one and zero, respectively.

Getting back to Eq.~\ref{eq:compute_similarity_weight}, we treat that two
back-propagation signals are similar, when their peak coefficient
is bigger than
the threshold, i.e.,  $a_{cc} > a_{th}$.
In this case, we assign a higher weight according to the relative time delay of
the length of the signal, $(\frac{L-l_{cc}}{L})$; i.e., we give the highest
weight when two signals are matched without any delay, under the assumption that those two signals are originated from the same sound source.

\Skip{
if it is
determined that two back-propagation signals are highly correlated because of
$(a_{cc} > a_{th})$; two back-propagation signals are caused by the same
source.
}


\section{RESULTS and DISCUSSION}
\label{sec:result_and_discussion}


In this section, we show how our approach accurately estimates the sound source
location by measuring distance errors between the ground truth and the
estimated position.
The yellow disk in Fig.~\ref{fig:summary} represents a 95\% confidence area for
the estimated source.  We also compare distance errors of our approach to the
prior work (RA-SSL) to demonstrate the effectiveness of our algorithm
considering the back-propagation signals; RA-SSL is the version of our approach without using the similarity of the back-propagation signals.

The hardware platform consists of Eigenmike, which is the 32-channel microphone array of the mh acoustics, and the i7 CPU computer.
For reconstructing indoor environments, we first collect a point cloud by using
Kinect v1 and then build a mesh map consisting of triangles from the point
cloud; the reflection coefficients are appropriately assigned to the triangles
by referring the reported values in \cite{schissler2018acoustic}.  

We report values of parameters used for our algorithm: $\alpha$ for controlling
the influence of each weight is $1$, the standard
deviation $\sigma_w$ of the Gaussian distribution function used for computing
the distance weight is 0.5 that is determined by the consideration of the size of the indoor environment (about one tenth of the room width 7m), and the threshold value $a_{th}$ for checking the
correlation between back-propagation signals is 0.15.  
We also show the results over different parameter values in
Sec.~\ref{sec:moving_sound_around_obs}.

We 
use 3840 samples for the separation signal, where
the sampling frequency is 48 kHz; 3840 audio samples (80~ms) are
a sufficient length for covering direct and first reflection signals as
indicated in \cite{chen2011time}.
\Skip{
The length of the back-propagation signals is 8192 audio samples because the
separation signals are resized to consider the time shifting of the
back-propagated signals based on the propagation delay.
}

\subsection{Benchmarks}

\begin{figure}[t]
	\centering
	\subfloat[The environment without the
	obstacle.]{\includegraphics[width=0.75\columnwidth]{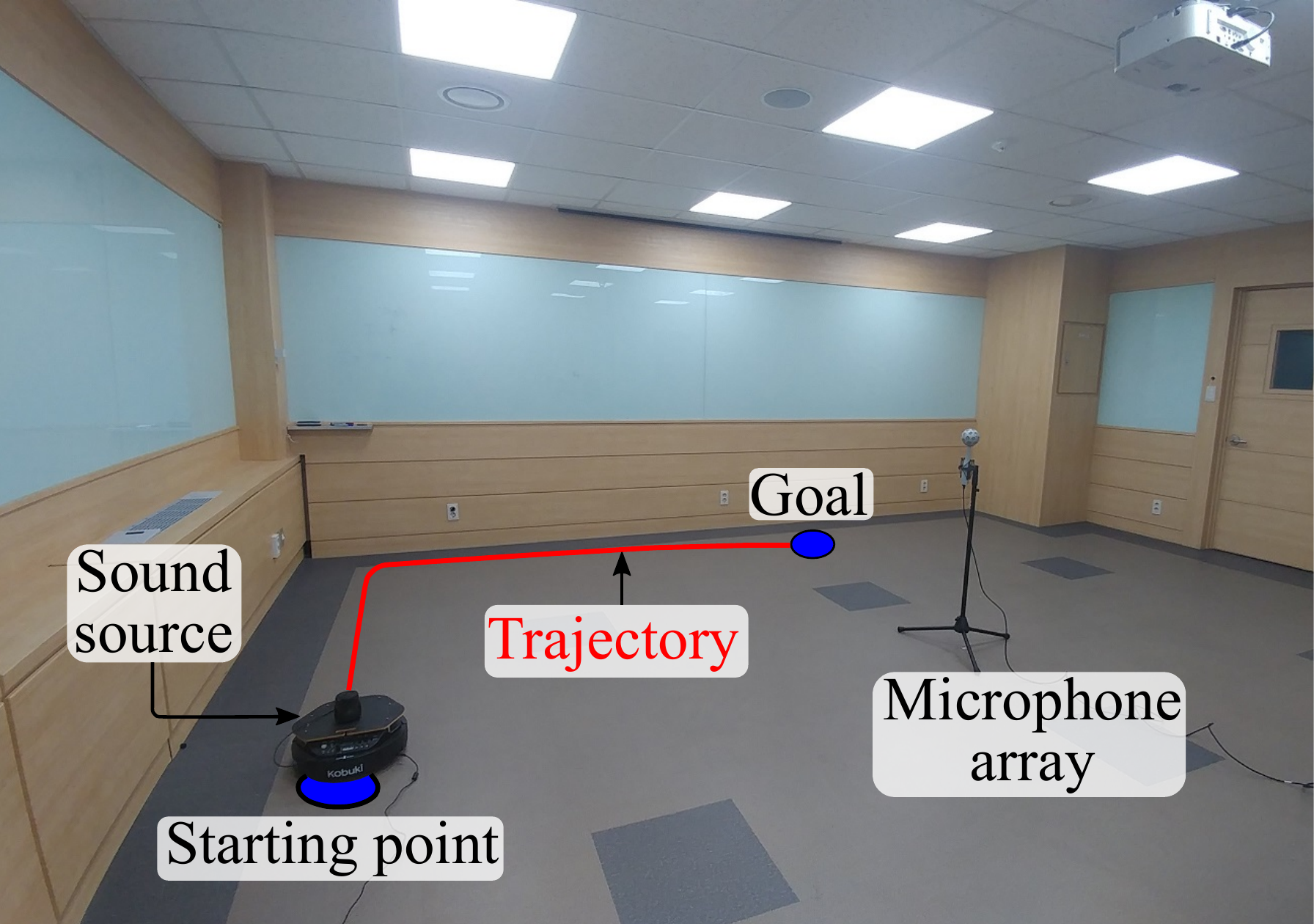}\label{fig:environment_wo_obs}}
	\
	\subfloat[The environment with the obstacle.]{\includegraphics[width=0.75\columnwidth]{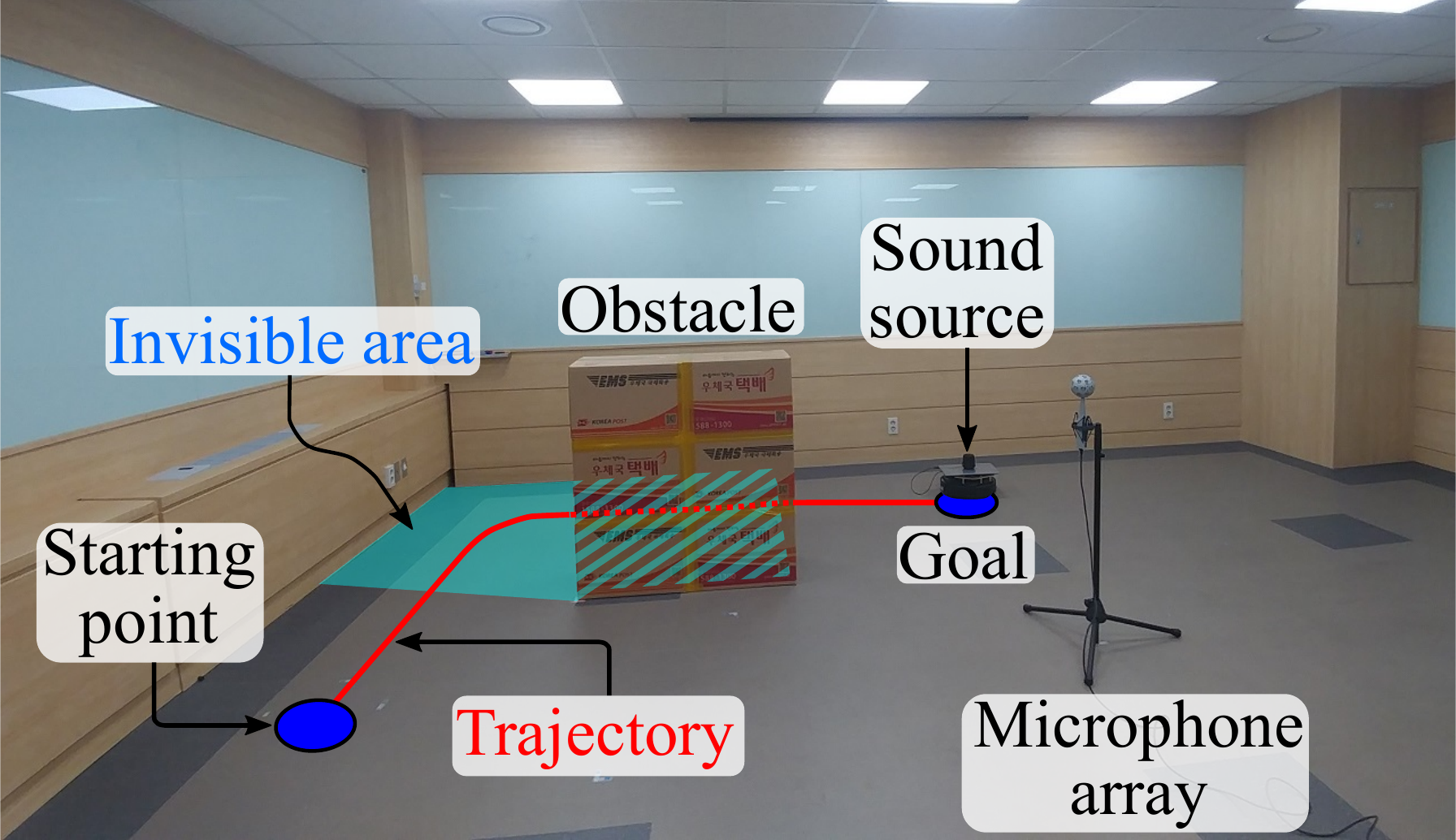}\label{fig:environment_w_obs}} \qquad
	\caption{
		The test environments w/ and w/o an
obstacle that can make the sound source non-line-of-sight one.
We use the clapping sound as the sound source.
\Skip{
		Where the sound source moves along with the red trajectory, 
		the microphone array collects the audio signals in (a).
		In (b), the obstacle cause the invisible area for the microphone array and the sound source becomes the non-line-of-sight state.
}
	}
	\vspace{-0.5em}
	\label{fig:environment}
\end{figure}

Different experiments were conducted in two scenes: the moving sound without and
with an obstacle.  In both environments (Fig.~\ref{fig:environment_wo_obs} and
Fig.~\ref{fig:environment_w_obs}), a robot equipped with an omni-directional
speaker moved along the red trajectory, and the 32-channel microphone array
recorded the audio signals, and these data are used for various tests with the
ground truth information on the sound source locations.  In
Fig.~\ref{fig:environment_w_obs}, we put an
obstacle made by paper boxes,
to cause the robot invisible along the robot's trajectory for the microphone
array; at the invisible area, the sound source becomes the
non-line-of-sight (NLOS) source.  

Handling the NLOS source was reported to be a quite difficult problem in
RA-SSL, because direct sound propagation paths are blocked by the obstacle and
we have to rely on indirect sound paths that are incoherent and sensitive to
noise.  
Furthermore, the number of indirect acoustic ray paths passing near the ground
truth is usually small, and thus the accuracy of the localization algorithm
tends to deteriorate.

Additionally, these scenes are not free from noise naturally occurring in a typical environment; they are exposed to noises,
as shown in Fig.~\ref{fig:visualize_noises}, since they are not controlled
scenes.  Noises can cause to trigger many
incoherent acoustic ray paths, hindering them to converge in a single location.

\begin{figure}[t]
	\centering
	\subfloat[Accuracy of moving sound w/o the obstacle
	(Fig.~\ref{fig:environment_wo_obs}).]{\includegraphics[width=0.75\columnwidth]{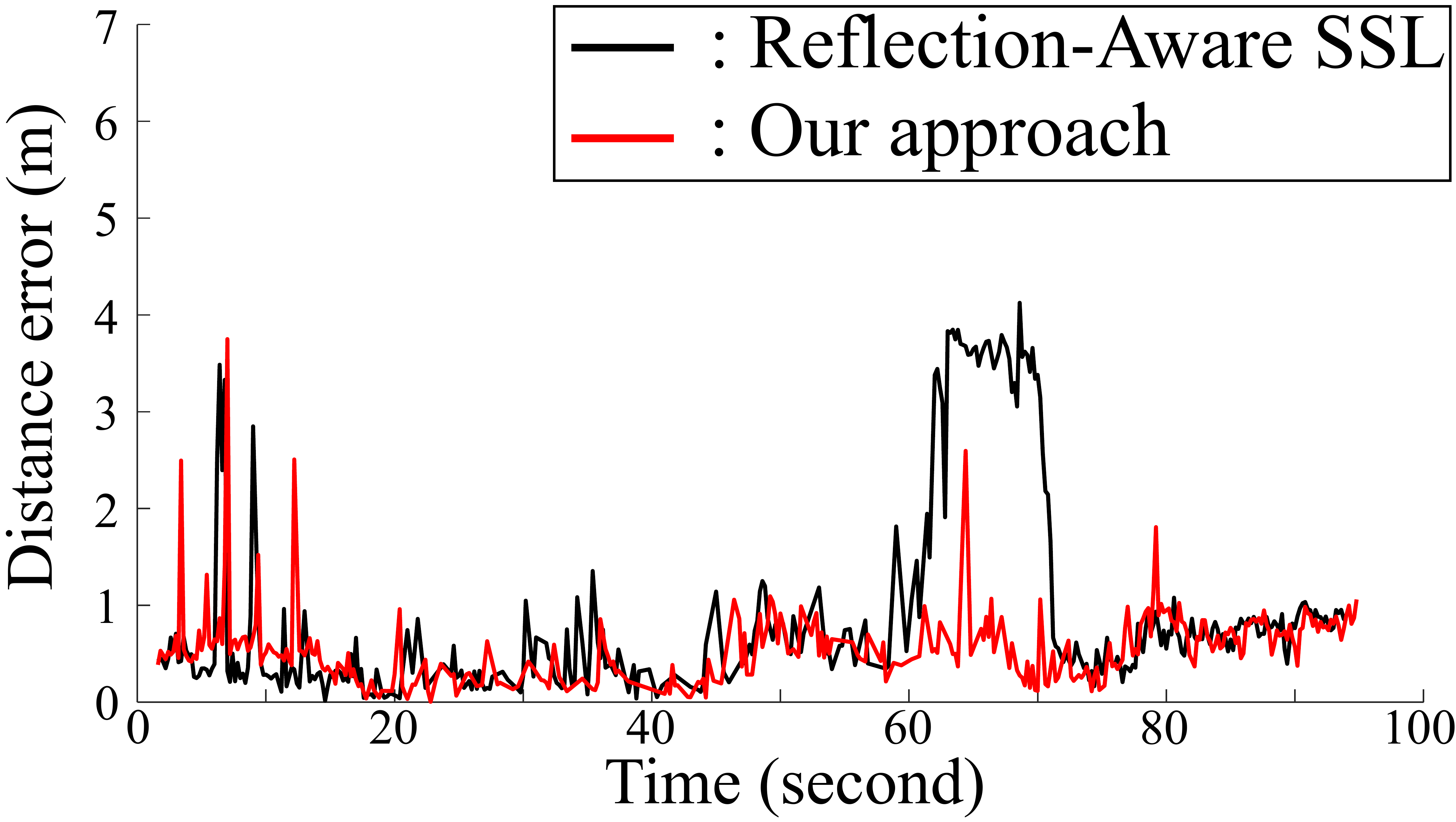}\label{fig:graph_wo_obs}}
	\
	\subfloat[Accuracy of moving sound w/ the obstacle (Fig.~\ref{fig:environment_w_obs}).]{\includegraphics[width=0.75\columnwidth]{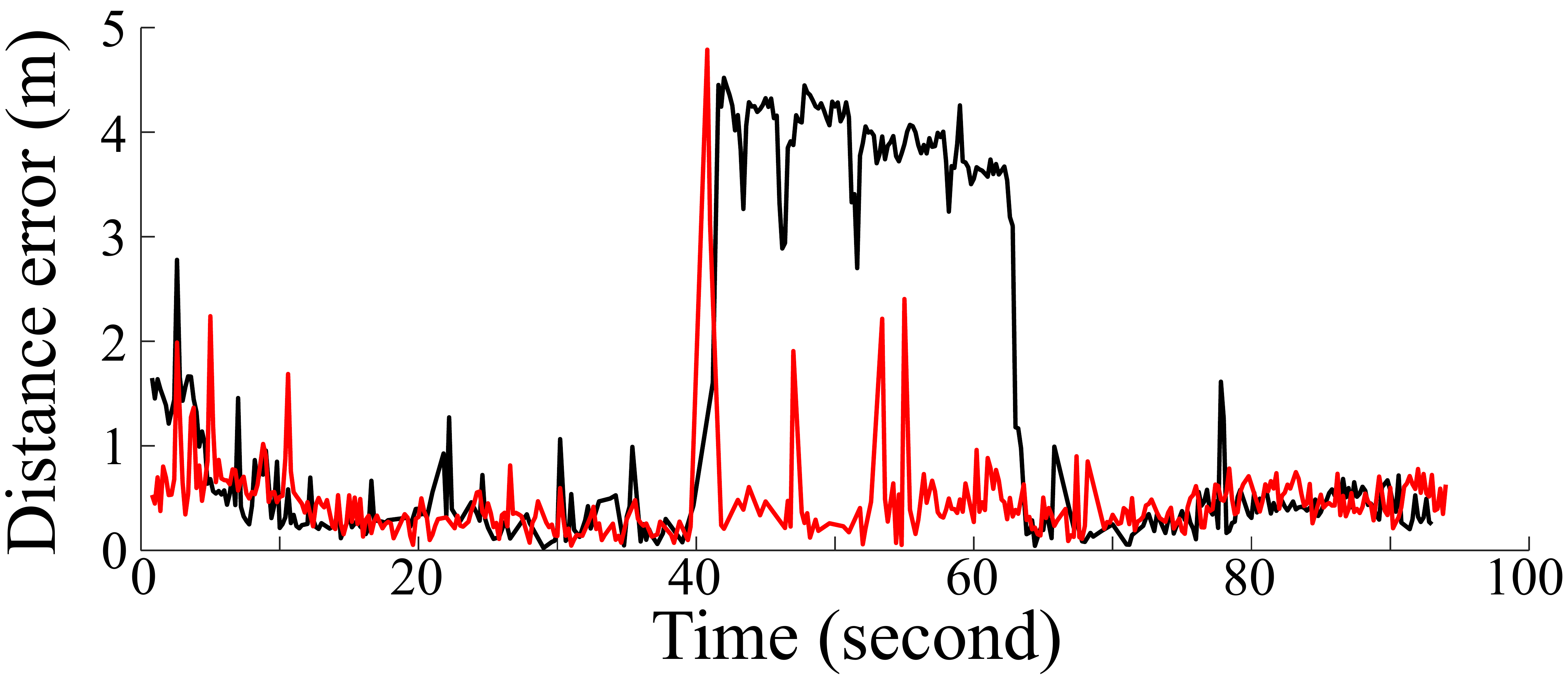}\label{fig:graph_w_obs}}\qquad
	\caption{
		\YOON{use thicker graph lines}
		The distance errors between the ground truth
		and the estimated source positions, where the black line is for
		the prior work (RA-SSL) and the red line is for our approach.
	}
	\vspace{-0.5em}
	\label{fig:graph}
\end{figure}

\begin{figure}[t]
	\centering
	\includegraphics[width=0.85\columnwidth]{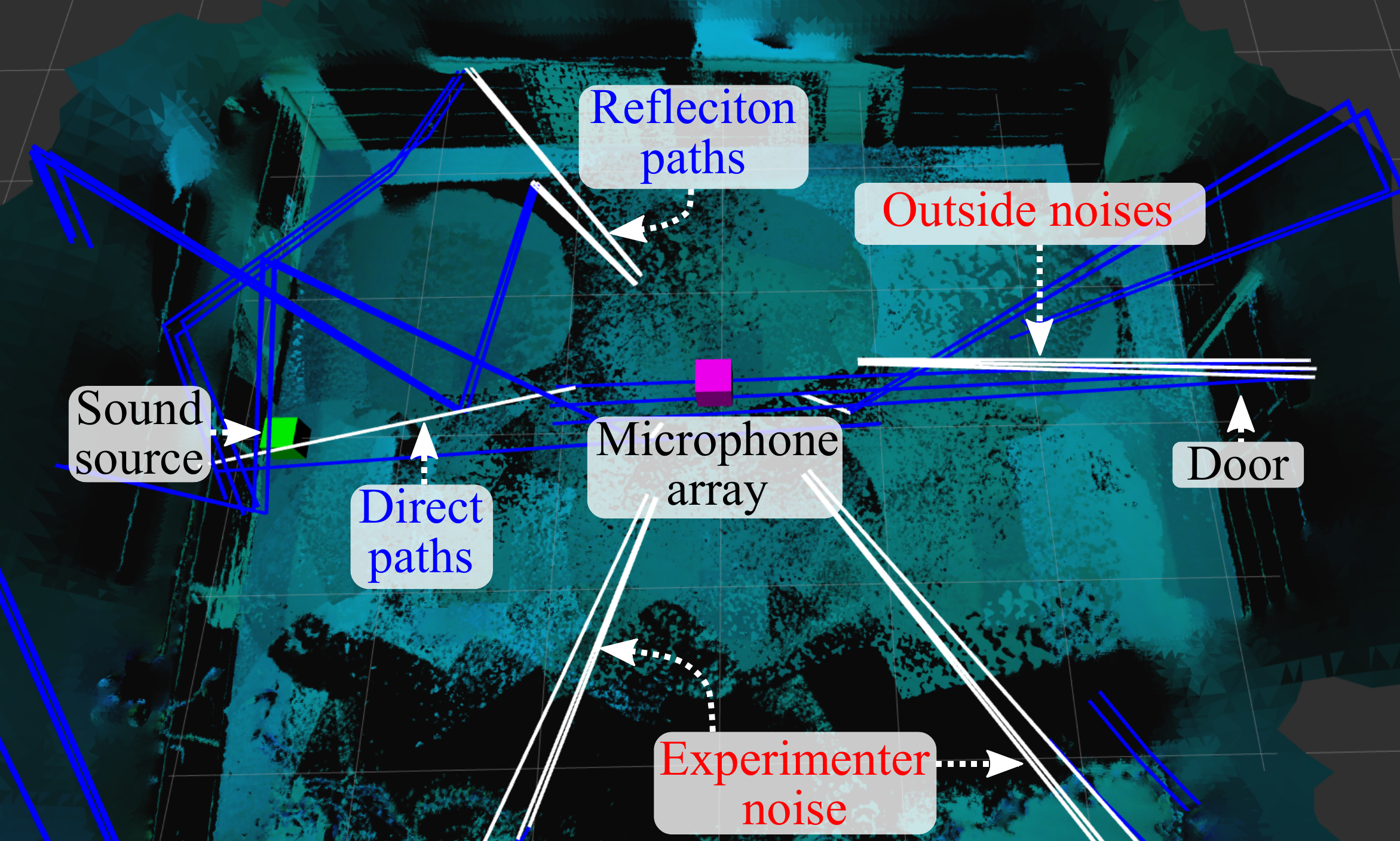}
	\caption{
		The test environment contains
many noises, some of which come from the outside and experimenters at the
bottom.
They generate incoherent acoustic rays, in addition to ones from the sound source.
\Skip{
are generated by the noise of experimenters and the noise sound that comes from
outside.  The sound source causes the acoustic ray paths of the reflection and
direct sound propagation paths.		
}
	}
	\vspace{-1.0em}
	\label{fig:visualize_noises}
\end{figure}

\subsection{A moving sound source}
\label{sec:moving_sound}

We first show how our approach has the advantage compared to RA-SSL in a simple
scene with a moving sound.  In Fig.~\ref{fig:graph_wo_obs}, the black and red
graphs denote the distance errors of RA-SSL and our approach, respectively.
The
average distance errors across  the whole test time are 0.9231m for RA-SSL and
0.5594m for our approach; the accuracy of the sound source localization is
improved about 65\% based on our approach.  

The experiment environments in Fig.~\ref{fig:environment} have many kinds of
noises, caused by experimenters and the outside, as shown in
Fig.~\ref{fig:visualize_noises}, and these noises generate acoustic rays that
do not help to localize the sound source.
Especially, when the moving sound source turns the corner from 60 s to 70 s, the accuracy
of RA-SSL deteriorates significantly.  In this case, 
RA-SSL
estimates source positions incorrectly near the noisy area
(Fig.~\ref{fig:visualize_noises}), while their signals from the noise source and from the ground truth source are different.
On the other hand, the red graph
shows that our method is robust even in this case, thank to considering the
back-propagation signals on estimated source locations; the similarity weight
improves the robustness of the source localization algorithm.


\subsection{A moving sound around an obstacle}
\label{sec:moving_sound_around_obs}

We now show results with the more challenging environment including an
obstacle between the source trajectory and the microphone array shown in
Fig.~\ref{fig:environment_w_obs}.  Fig.~\ref{fig:graph_w_obs} shows graphs
of  the distance errors of RA-SSL and our approach.
The average distance errors of RA-SSL and our
approach are 1.4919m and 0.4623m, respectively.  Especially, where the sound
source is in the NLOS state from 40 to 70 seconds, the accuracy of RA-SSL
decreases drastically, because blocking the direct sound propagation paths
makes the convergence of acoustic rays weak near the ground truth.
On the other hand, even in this challenging case, we get a stable result, 
220\% improvement compared to RA-SSL, by considering similarity
between back-propagation signals of indirect acoustic paths.

To analyze effects of varying values of $\alpha$, $\sigma_w$, and $a_{th}$, we
measure the average distance errors over various parameter values.
Table~\ref{table:results_parameter} shows that our algorithm is robust to
changes of parameters, where values of $\alpha$, $\sigma_w$, and $a_{th}$ vary
from 0.5 to 1.5, from 0.3 to 0.7, and from 0.1 to 0.15, respectively.  However,
if the value of $a_{th}$, the threshold for checking the peak
coefficient, becomes 0.17, the accuracy dramatically decreases.
This is mainly because coefficients of pairs of back-propagation signals
originated by the same source have values from 0.17 to 0.23.
As a  result, when $a_{th}$ becomes too large, equal to or bigger than 0.17, we
even filter out similar signals, and this enforces our approach to fall back to
behave like the prior method RA-SSL.  Nonetheless, our approach even in this
case 
outperforms the prior method; RA-SSL's average error is
1.4919m.
\Skip{
the number of back-propagation signals, considered in the similarity weight,
decreases because of the condition, $a_{th}<a_{cc}$, in the similarity weight
and the accuracy result
becomes same with RA-SSL, only considering the distance weight.
}

\begin{table}	
	\caption{The average distance errors over various parameter values}
	\begin{center}
		\begin{tabular}{clcccc}
			\Xhline{3\arrayrulewidth}
			$\alpha$ 		&	0.5		&	0.75	&	1.0		&	1.25	&	1.5		\\
			\hline
			$\sigma_w=0.5$, $a_{th}=0.15$	&	0.55m	&	0.52m	&	0.46m	&	0.63m	&	0.6m	\\
			\Xhline{3\arrayrulewidth}
			$\sigma_w$ 		&	0.3		&	0.4		&	0.5		&	0.6		&	0.7		\\
			\hline
			$\alpha=1.0$, $a_{th}=0.15$	&	0.44m	&	0.48m	&	0.46m	&	0.57m	&	0.65m	\\
			\Xhline{3\arrayrulewidth}
			$a_{th}$ 		&	0.1		&	0.125	&	0.15	&	0.175	&	0.2		\\
			\hline
			$\sigma_w=0.5$, $\alpha=1.0$	&	0.52m	&	0.47m	&	0.46m	&	1.12m	&	1.3m	\\
			\Xhline{3\arrayrulewidth}
		\end{tabular}
	\end{center}
	\vspace{-1.5em}
	\label{table:results_parameter}
\end{table}

\subsection{Analysis of back-propagation signals}

\begin{figure}[t]
	\centering
	\includegraphics[width=0.9\columnwidth]{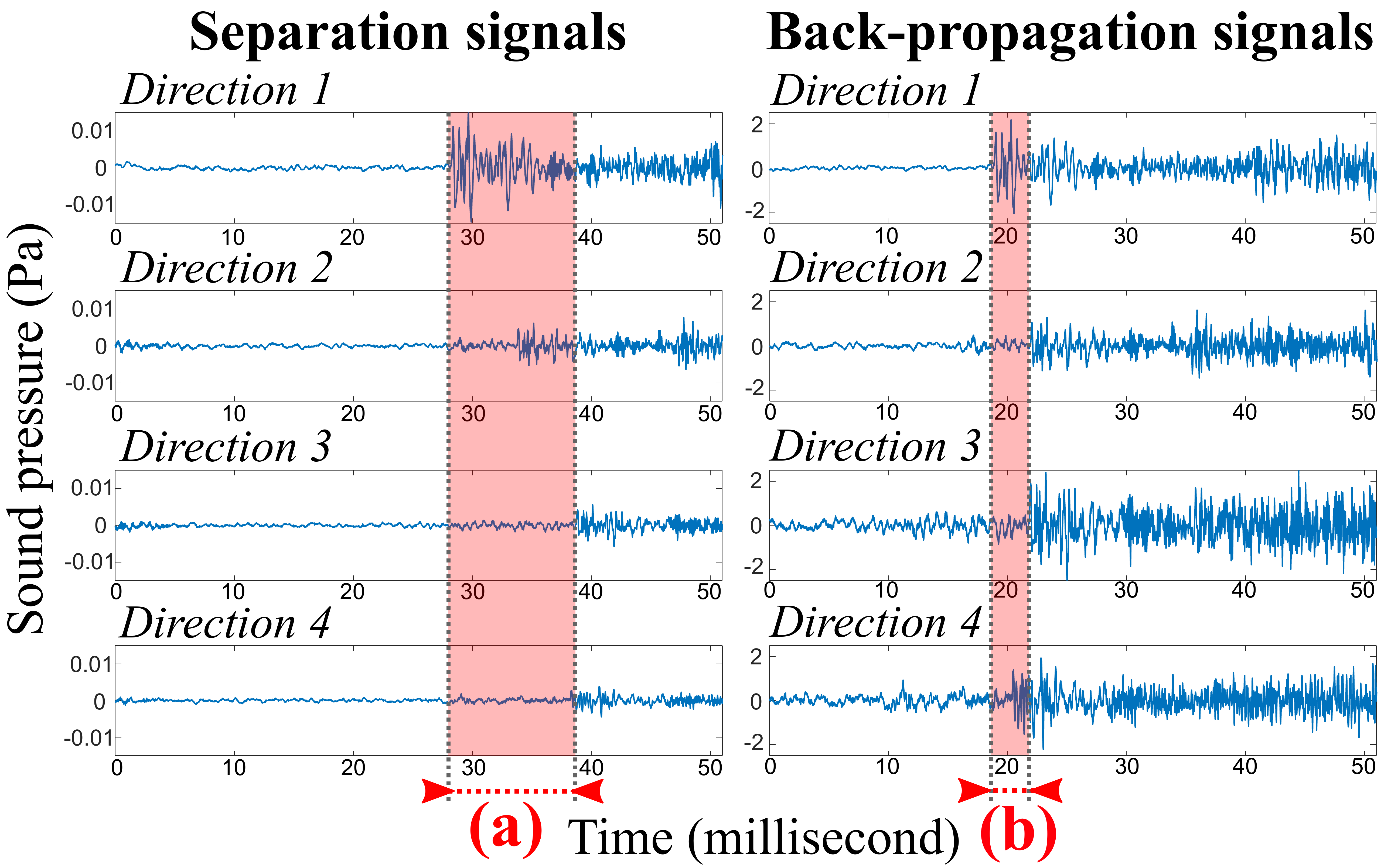}
	\caption{
		On the left, we show separation
		signals heard from different incoming directions of the same 
		sound source, while the right side shows their corresponding,
		back-propagation signals.  The widths of red rectangles, (a)
		and (b), represent the time differences of the separation and
		back-propagation signals.  The time difference (b) of the
		back-propagation signals is smaller than the separation signals
		(a), indicating
		that the back-propagation signals are more
		similar each other compared to 
	the separation
		signals. 
	}
	\vspace{-1em}
	\label{fig:analysis_back_propagation_signals}
\end{figure}

Let us see how back-propagation signals have positive effects on the 3D sound
source localization.  
Fig.~\ref{fig:analysis_back_propagation_signals} shows
four separation signals observed from different  incoming directions of the same sound source.
On the right side of the figure, we  also show four back-propagation signals
generated from those observed separation signals at a location of the ground
truth.  The width of the red rectangle (a) shown on the left side of the figure
indicates the time difference, caused by the distance difference of sound
propagation paths, of separation signals.  After computing the back-propagation
signals from those separation signals, we observe that the time difference,
denoted by the width of the red rectangle (b), of back-propagation signals was
reduced.  The reduction of the time difference in (b) compared to (a) can be
interpreted that the back-propagation signals are restored better, since they
are all originated from the same sound source.


We also measure cross-correlation between signals.
When every pair of separation signals in
Fig.~\ref{fig:analysis_back_propagation_signals} are analyzed by the
cross-correlation operation, the average of peak coefficients and peak
coefficient delays are 0.2183 and 280 samples, respectively.  For the
back-propagation signals, the average of peak coefficients and peak coefficient
delays are 0.2245 and 35 samples.  These values indicate that the computed
back-propagation signals are restored in a way that those signals are similar to each other.
Note that a higher peak coefficient indicates
more correlated signals, and
the peak coefficient delays close to zero represents that signals are well-aligned in time.
\Skip{
Because the condition, $a_{cc} > a_{th}$, is satisfied where $a_{th}$ is 0.15 and the peak coefficient delays, $l_{cc}$, are small enough, these back-propagation signals have a high similarity weight $w_s$.
}

\section{Limitations and Conclusion} 
\label{sec:conclusion}

We have presented a novel sound source localization algorithm using
back-propagation signals.  After estimating propagation paths of the sound by
generating acoustic ray paths, the back-propagation signal virtually computed
at a specific point on the acoustic ray path 
is considered.  
We utilize those back-propagation signals of different acoustic paths for
robustly identifying the converging region of the source, even in
environments with noises and an obstacle.

\Skip{
 design and
utilize a novel weight of particles of the
Monte-Carlo localization algorithm for improving a localization accuracy.
}

While we have demonstrated benefits of our approach, it has several limitations
and opens up many interesting future directions.
In Fig.~\ref{fig:graph}, in some cases, our accuracy is lower than the prior work because of cropping a specific length of an audio signal from a measured signal at the microphone every fixed cycle; the cropped signal may not be able to collect enough audio signal at the beginning of the sound.
We plan to deal with this problem by cropping and processing a meaningful audio signal from a measured signal at the microphone.
\YOON{In Fig. 8, in some cases, our accuracy is lower than the prior work. Can you explain why it happens and how we can address it here briefly?}
  Currently, real-time
computation of our method is not ensured, due to the premature implementation of our current proof-of-concept system;
the beamforming module runs in Matlab and the cross-correlation operation
performed in pairs of available acoustic ray paths (e.g., 8 paths on average in
our tests) runs serially.  We plan to address this issue by re-implementing the
beamforming module in C++ and
designing the cross-correlation operations  in a parallel manner.

Other sound propagation phenomena that are frequently observed 
at low frequencies or in the low frequency region 
such as scattering and diffraction are not handled yet.
Fortunately, a ray tracing based approach supporting the diffraction
effect is recently proposed~\cite{an2018diffraction}, and can be adopted for our method.
The acoustic material properties such
as reflection coefficients of triangles of objects are not
automatically assigned, and 
some of deep learning approaches showing promising results can be employed to solve this problem~\cite{schissler2018acoustic}.



\bibliographystyle{plainnat}

\end{document}